\documentclass{article}

\usepackage{arxiv}

\usepackage[utf8]{inputenc} 
\usepackage[T1]{fontenc}    
\usepackage{hyperref}       
\usepackage{url}            
\usepackage{booktabs}       
\usepackage{amsfonts}       
\usepackage{nicefrac}       
\usepackage{microtype}      
\usepackage{lipsum}		
\usepackage{graphicx}
\usepackage{doi}
\usepackage{amsmath,amsfonts}
\usepackage{algorithmic}
\usepackage{algorithm}
\usepackage{array}
\usepackage[caption=false,font=normalsize,labelfont=sf,textfont=sf]{subfig}
\usepackage{textcomp}
\usepackage{stfloats}
\usepackage{url}
\usepackage{verbatim}
\usepackage{graphicx}
\usepackage{braket}
\usepackage{xcolor}

\usepackage[backend=bibtex, style=ieee]{biblatex}
\title{Distribution of GHz sequential Time-bin
Entanglement in a Metropolitan Fiber Network}

\author{
\hspace{1mm}Martin Achleitner\\
	Digital Safety and Security\\
	AIT - Austrian Institute of Technology\\
	1210 Vienna, Austria\\
	\And
    \hspace{1mm}Alessandro Trenti\\
	Digital Safety and Security\\
	AIT - Austrian Institute of Technology\\
	1210 Vienna, Austria\\
    \And
    \hspace{1mm}Philip Walther \\
    Faculty of Physics\\
	University of Vienna\\
	1090 Wien, Austria\\
    \And
    \hspace{1mm}Hannes Hübel \\
	Digital Safety and Security\\
	AIT - Austrian Institute of Technology\\
	1210 Vienna, Austria\\
    }

\renewcommand{\shorttitle}{\textit{arXiv} Template}

\hypersetup{
pdftitle={A template for the arxiv style},
pdfsubject={q-bio.NC, q-bio.QM},
pdfauthor={Martin Achleitner, Alessandro Trenti, Philip Walther, Hannes Hübel},
pdfkeywords={Time-bin Entanglement, Fiber network},
}

\begin{document}
\maketitle

\begin{abstract}
Efficient generation and high-quality distribution of entanglement is becoming increasingly more relevant in the field of quantum technologies, with important applications such as multiparty computation as well as quantum key distribution (QKD) on the rise. Quantum communication protocols based on entanglement offer an inherent quantum based randomness for key generation and provide in general higher security compared to prepare and measure implementations. Moreover, the future quantum internet will also be based on the distribution of entanglement for securely connecting quantum computers in a network.

In this work we show the feasibility of using sequential time-bin entangled states for quantum key distribution in metropolitan networks using off-the-shelf components. The time-bin encoding ensures high fidelity distribution robust against random polarisation fluctuations occuring in optical fibers. Modulated laser pulses in the GHz frequency range are used to generate time-bin entangled photon pairs. The entangled photons are then sent over an about 30km long (9.5dB loss) fiber link within the Vienna fiber network, showing high degree of distributed entanglement with a measured 93\% quantum visibility.
\end{abstract}

\section{Introduction}

{I}{n} the pool of QKD protocols there are two major categories, one is prepare-and-measure and the other entanglement based protocols \cite{bennett_quantum_1992, bennett_quantum_2014}. 
Prepare-and-measure protocols classically encode information on degrees of freedom such as polarisation of single photons, while entanglement based protocols leverage their inherent quantum randomness for key generation and pair correlation for communication. Prepare-and-measure protocols are usually easier to implement, however they need a source of unpredictable randomness for key generation. 
While already offering a high degree of security they are still vulnerable to certain attacks. Entanglement based QKD protocols on the other hand are typically relying on complex generation setups but benefit from their inherent randomness and photon pair correlation providing a higher security for key distribution by relaxing trust assumptions on the system.\\ 
In many entanglement based QKD protocols, polarisation is used as the entangled degree of freedom, since it is easily generated and manipulated. For distribution over long fiber links however polarisation is hard to keep stable due to the inherent birefringence of the fiber. Other entanglement generation schemes like time-bin entanglement \cite{de_riedmatten_tailoring_2004,thiel_time-bin_2024, anwar_entangled_2021, marcikic_time-bin_2002} where photon pairs are correlated across different time-bins offer a promising solution for this. Protocols based on time-bin entanglement do not rely on a stable polarisation for entanglement distribution, their entangled degree of freedom is the uncertainty in generation time. However time-bin entanglement sources typically rely on complex interferometric setups requiring challenging custom-designed solutions \cite{fitzke_scalable_2022}.\\

Since a future quantum internet connecting nodes containing quantum computers will be offering not only secure QKD connections but also beyond QKD applications like multiparty computation and teleportation, entanglement generation and distribution will be an increasingly important resource. Our goal is to generate high quality and high rate entanglement enabling distribution over long distances. For this work we implemented a sequential time-bin entangled source which was used to demonstrate entanglement distribution in a metropolitan network.
In this scheme the probability wave of downconverted photon pairs in two consecutive laser pulses are overlapping in a delay line interferometer to generate ambiguity of the originating time-bin of the pairs. In this work we built and investigated such a scheme including an off-the-shelf Mach-Zehnder interferometer (MZI) delay line.
Laser pulses were generated in the GHz range and sent through a spontaneous parametric downconversion crystal to generate photon pairs.\\
The entangled states were then sent via a fiber link of around 30km through the city of Vienna to show the feasibility of entanglement distribution in a metropolitan fiber network. The entangled pairs were finally measured with superconducting nanowire single photon detectors at the University of Vienna, showing a high degree of entanglement. \\

This paper is divided into the following sections: section 2 describes the entanglement source together with the MZI delay line for entanglement analysis. In section 3 preliminary characterization measurements like coincidence to accidental ratio (CAR) and the SPDC output spectrum are reported. A theoretical examination of the sequential time-bin entanglement scheme is performed in section 4. Section 5 discusses the results and challenges from entanglement distribution performed through the inner city of Vienna. In section 6 we conclude the paper with a summary of the achieved results and future plans.

\section{Setup}
\begin{figure*}[t]
\centering
\includegraphics[width=0.9\linewidth]{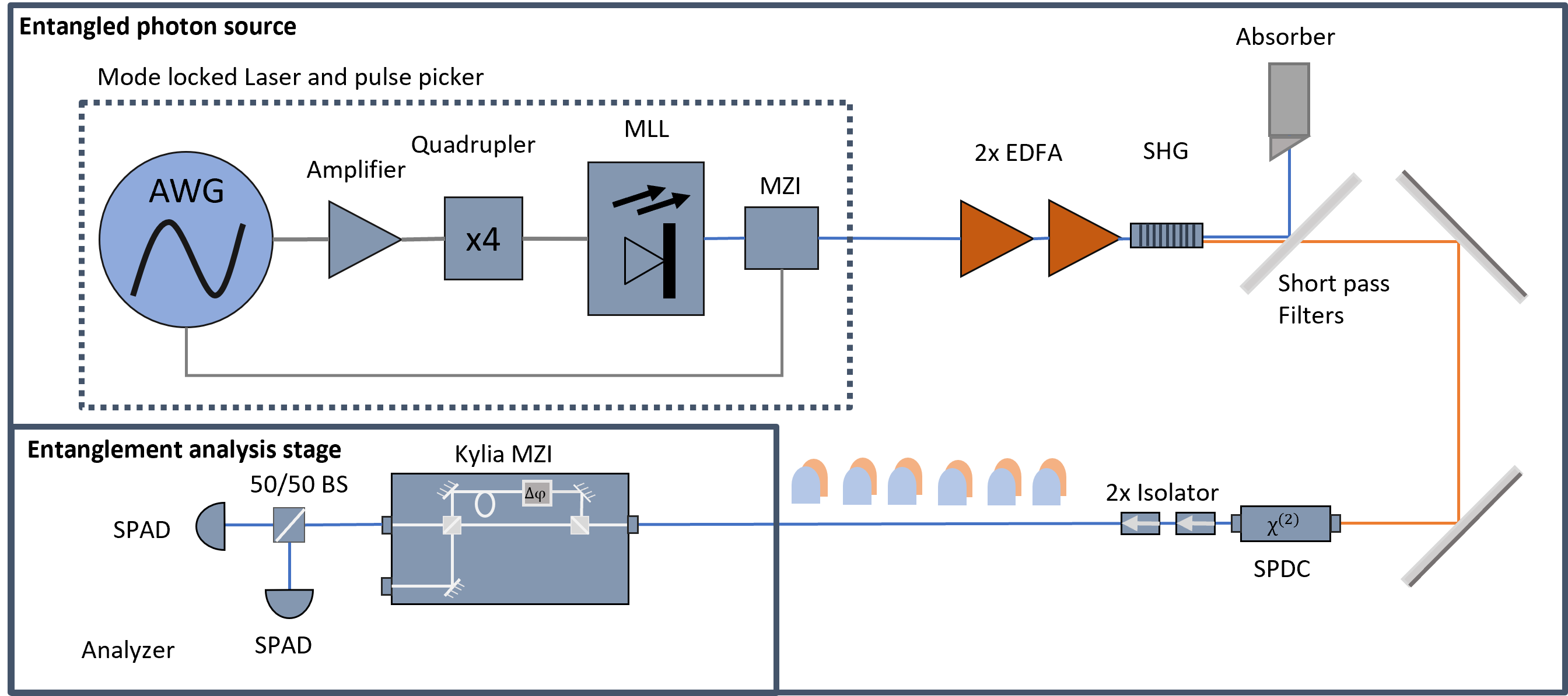}

\caption{The entanglement source is comprised of the pump setup with a laser (MLL) (1554nm), pulsegenerator(AWG), two EDFAs and a SHG stage that sends the generated light at 777nm after filtering by a dichroic mirror and an absorber to the pair generation stage. This consists of a type II SPDC crystal and 2x isolators (generated photon pairs are shown as orange and blue pulses). The time-bin entangled photon pairs are analyzed via a delay line Mach-Zehnder interferometer (MZI) and a 50/50 beamsplitter setup and measured by single photon avalanche detectors.}
\label{fig:Setup}
\end{figure*}

The  setup for generating and detecting time-bin entangled photon states is shown in Figure \ref{fig:Setup}. It is comprised of a pump setup, a photon pair generation stage and an entanglement analysis part.
For high-rate and short pulse generation a mode locked laser (MLL) with an additional pulse picker was used. The MLL is based on a multi-section semi-insulating planar buried heterostructure multiple quantum well mode locked laser diode based on GaInAsP/InP (manufactured by u2t photonics) . It includes an integrated saturable absorber (SA) and a distributed Bragg reflector (DBR) grating. The laser is mounted on a sub mount for temperature stabilization. The nominal repetition rate of the laser is 42.657GHz with a 2.5ps pulse width.
To lock the MLL repetition frequency to a master clock we employed an arbitrar waveform generator (AWG) in combination with a quadrupler to drive the absorber section of the MLL.
Since the native repetition rate at 42.657GHz is to high for time resolved measurements of time-bins (typical detector jitter is in the range of 100ps), an optical downsampling was implemented. For this a pulse picker was synchronized with the AWG to down-sample the rate to a more manageable frequency in the range of 1-10GHz \cite{zeiger_ps-pulse_2019}.\\

The light pulses are then amplified by two erbium doped fiber amplifiers (EDFA) to reach an input power of around 30dBm for the second harmonic generation (SHG) crystal to generate the visible pump beam (777nm) for the subsequent spontaneous parametric downconversion (SPDC) pair-generation stage. The SHG crystal is a MgO 5mm long periodically poled LiNb bulk crystal (manufactered by Covesion). SHG is phase-matched in the domains with 19.40$\mu$m or 19.10$\mu$m poling periods at temperatures around 33°- 34°C and 103°- 104°C respectively. 

Two short-pass filters are used after the SHG to filter out any remaining pump photons at 1554nm, providing a total rejection higher than 100dB. 
After optimization, the in-coupled power at 777nm is as high as 20mW. The visible pump power is then injected into the SPDC stage where the photon pairs at 1554nm are generated. The SPDC crystal is of type-II fiber pigtailed waveguide-based ppKTP (AdvR), which generates degenerate pairs of photons with orthogonal polarisation. The length of the crystal is 10mm. The poling period in KTP waveguides is estimated to be about 100$\mu$m, which depends on the properties of the bulk material of the chip and the properties of the specific waveguide.
The input and output coupling losses are estimated to be about ~50\%. The time slots potentially containing the generated photon pairs are visualized in figure \ref{fig:Setup} as orange and blue pulses.\\

After the photon pair generation the pairs are injected into the entanglement analysis stage where the quality of entanglement can be characterized. The entanglement stage is comprised of a Mach-zehnder delay line interferometer (MZI), the MZI module is a commercially available product manufactured by ixBlue (Kylia), a subsequent 50/50 beamsplitter (BS) was used for coincidence measurements. One output of the MZI is connected to the 50/50 BS with its two outputs connected to two single photon avalanche detectors.

\section{Characterization measurements}
For generating entangled states with a high repetition rate, the mode locked laser with a natural repetition rate of up to 42GHz was tested. With the help of the pulsepicker it was possible to downscale the pulse frequency of the MLL to 1GHz and 10GHz. 

Singles and coincidences were measured with two InGaAS Single-Photon Avalanche Diodes (SPADs, MPD-IR). The SPADs had a detection efficiency of about 8\%, dark-count rate of 850Hz, 150ps jitter and operated at a deadtime of 25µs for limiting afterpulsing. The TTL outputs for the SPADs are fed into a time-tagging module (TTM8000, AIT) which processes the data and outputs the singles and coincidence events in real time. The SPDC brightness was measured to be 7.2MHz/mW.

\begin{figure}[t]
\centering
\includegraphics[width=0.99\linewidth]{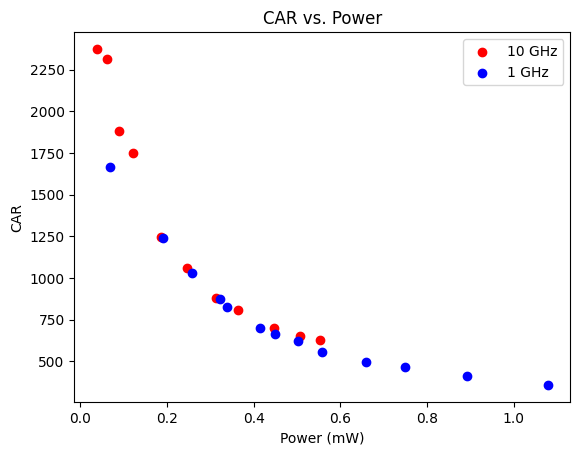}

\caption{Coincidences vs accidentals ratio visualized for different laser input powers, for 1Ghz and 10 GHz repetition rate.}
\label{fig:CAR}
\end{figure}
\begin{figure}[t]
\centering
\includegraphics[width=0.99\linewidth]{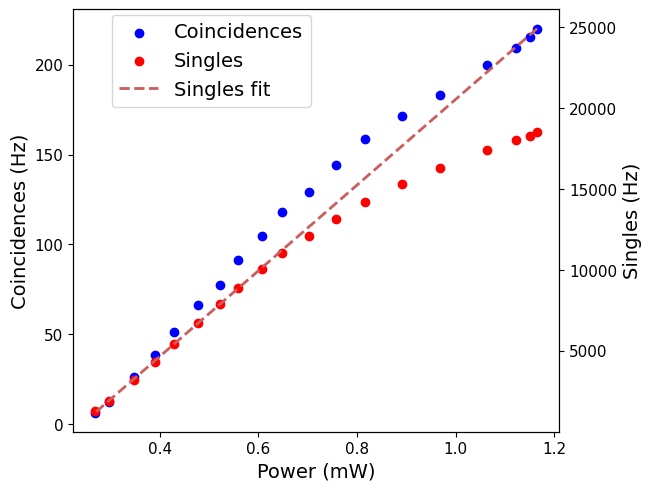}

\caption{The linear increase of coincidence count rate (blue) as well as single count rate (red) with growing power for a repetition rate of 1GHz is shown. An efficiency decrease of the detectors for higher power due to saturation effects is visible starting from around 0.7mW. The singles rate is fitted for values before 0.7mW (red dashed line) after 0.7mW the decreased linearity of single photon detections can be seen.}
\label{fig:CoincPower}
\end{figure}

The coincidence rate and the coincidences to accidental ratio (CAR) were measured for pulse repetition rates of the MLL of 1GHz and 10GHz. The CAR is calculated by taking the ratio of the coincidences within a coincidence window of 350ps (limited by the detection jitter) to the average of the background coincidences (accidentals) within the same time window.
In figure \ref{fig:CAR} the CAR vs. input laser power (P) is visualized, the 1/P relation of the CAR is clearly visible in the power range 0.1 to 0.3 mW which is explained by multipair emission of the SPDC at higher powers \cite{de_riedmatten_tailoring_2004}. For higher input powers the SPADs enter a nonlinear regime which modifies the slope. In the low pump power regime, the CAR reaches a maximum and it is eventually limited by the SPAD dark counts.

The generation probability of a photon pair per pulse for an average input power of 0.5mW is 0.18\% for 1GHz and 0.018\% for 10GHz. 
Both probabilities are well below the 10\% threshold for minimizing the contribution of multi pairs in the emitted photon pairs \cite{jin_efficient_2014}. As it is possible to see from Fig \ref{fig:CAR}, due to the low SPDC generation probability per pulse (since we are well below the 10\% threshold), the CAR does not substantially differ between 1 and 10GHz.

In the low pump power regime the coincidence rate and the singles are expected to increase linearly with increase in input power \cite{mancinelli_mid-infrared_2017, jin_efficient_2014}, which is shown in figure \ref{fig:CoincPower}. 

The drop in efficiency for higher input powers, due to saturation effects of the SPADs (deadtime: 25µs) is shown in the decreasing slope from after around 0.7mW. The singles were linearly fitted until this point, shown as a dark blue line in figure \ref{fig:CoincPower}. 
From the linear fit taking into account all the loss factors from generation to detection the SPDC efficiency was estimated to be $9.2\cdot 10^{-10}$. 

\begin{figure}[h!]
\centering
\includegraphics[width=0.99\linewidth]{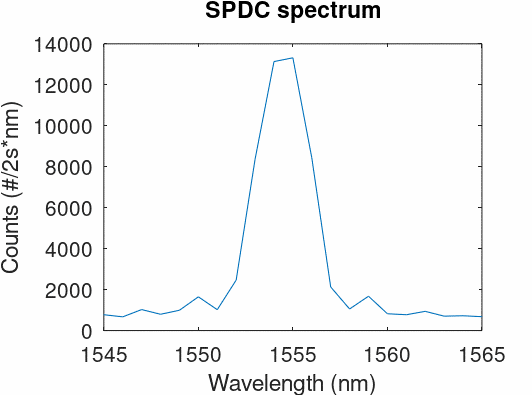}

\caption{Spectrum of the SPDC output measured with OSA operated as monochromator and SPAD. The spectrum is centered around the degenerate wavelength of around 1554nm with a bandwidth of approximately 3.4nm.}
\label{fig:SPDCspectrum}
\end{figure}

The spectrum of the SPDC photons is shown in Fig. \ref{fig:SPDCspectrum}. It is measured with an optical spectrum analyser (OSA) in combination with an external single photon avalanche detector (SPAD). With a SPAD as a measuring device and the OSA operating as a monochromator it is possible to get a single photon resolved SPDC spectrum. As expected the spectrum is centered around the degenerate wavelength of approximately 1554nm with a bandwidth of around 3.4nm.

The MZI delay line was first tested classically with a modulated butterfly laser at 1GHz repetition rate and a pulse duration of 150ps. The MZI is a commercial device from the company iXBlue and is polarisation independent which eases the manipulation of the SPDC photons.
The Kylia MZI has a free spectral range of 1GHz and an insertion loss of 1.6dB. To characterize the extinction ratio of the MZI, the phase in the delay arm was tuned, which works by thermal tuning based on Joule and thermo-optic effects.

\begin{figure}[h!]
\centering
\includegraphics[width=1.08\linewidth]{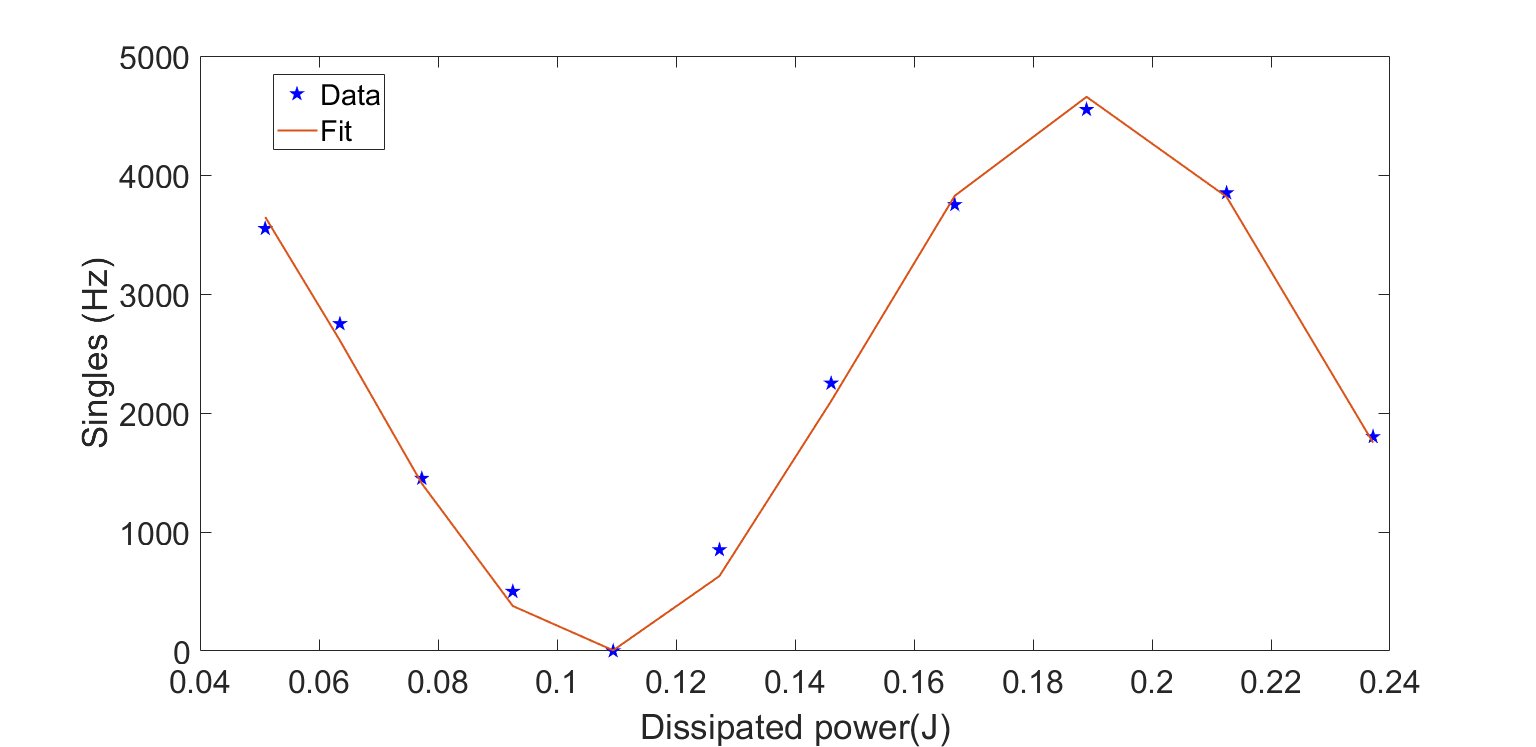}

\caption{Classical interference measurement of the MZI. 
The red curve is a cosine fit of the experimental data, reported as blue stars.}
\label{fig:classicalInterference}
\end{figure}

In Fig. \ref{fig:classicalInterference} classical interference was measured (blue stars). The experimental data are well fitted by a sinusoidal curve with a resulting visibility of around 99.77\% inferred from the fit function.

\section{Sequential Time-bin Entanglement}
To generate sequential time-bin entangled states, a pump pulse train gets fed into a type II SPDC crystal, which generates with a certain probability, orthogonally polarised photon pairs. The resulting state after the SPDC can be written as a d-dimensional quantum state with probability amplitudes $c_j$ and phase $\phi_j$ \cite{de_riedmatten_tailoring_2004}
\begin{equation}
    \ket{\psi}=\sum^d_{j=0} c_j e^{i\cdot \phi_j} \ket{j_s,j_i}
    \label{eq:spdcstate}
\end{equation}

where j labels the consecutive time-bins containing signal and idler photons.
In the 2-dimensional time-bin entangled scheme the generated photon pulses are sent to an unbalanced MZI with a free spectral range matched with the repetition rate of the pulsed laser. At the MZI the photon pairs have different distribution possibilities, either they split up and the idler photon takes the delayed path and the signal photon the short path or vice versa, or the photons take the same path through the interferometer. In the latter case the photon pairs are correlated in sequential time-bins, meaning that there is an uncertainty if the pair took the short path and is coming from the current time-bin or if it took the long path and originated from the previous time-bin. 
The photon pair coming from the same time-bin will transform in the MZI in the following way \cite{riedmatten_creating_2002}:
\begin{equation}
    \ket{j,j}\rightarrow \ket{j,j}+e^{2i\delta}\ket{j+1,j+1}+e^{i\delta}\ket{j,j+1}+e^{i\delta}\ket{j+1,j}
    \label{eq:timebintransform}
\end{equation}

where we dropped the signal and idler subscripts for better readability.

\begin{figure}[h!]
\centering
\includegraphics[width=0.5\linewidth]{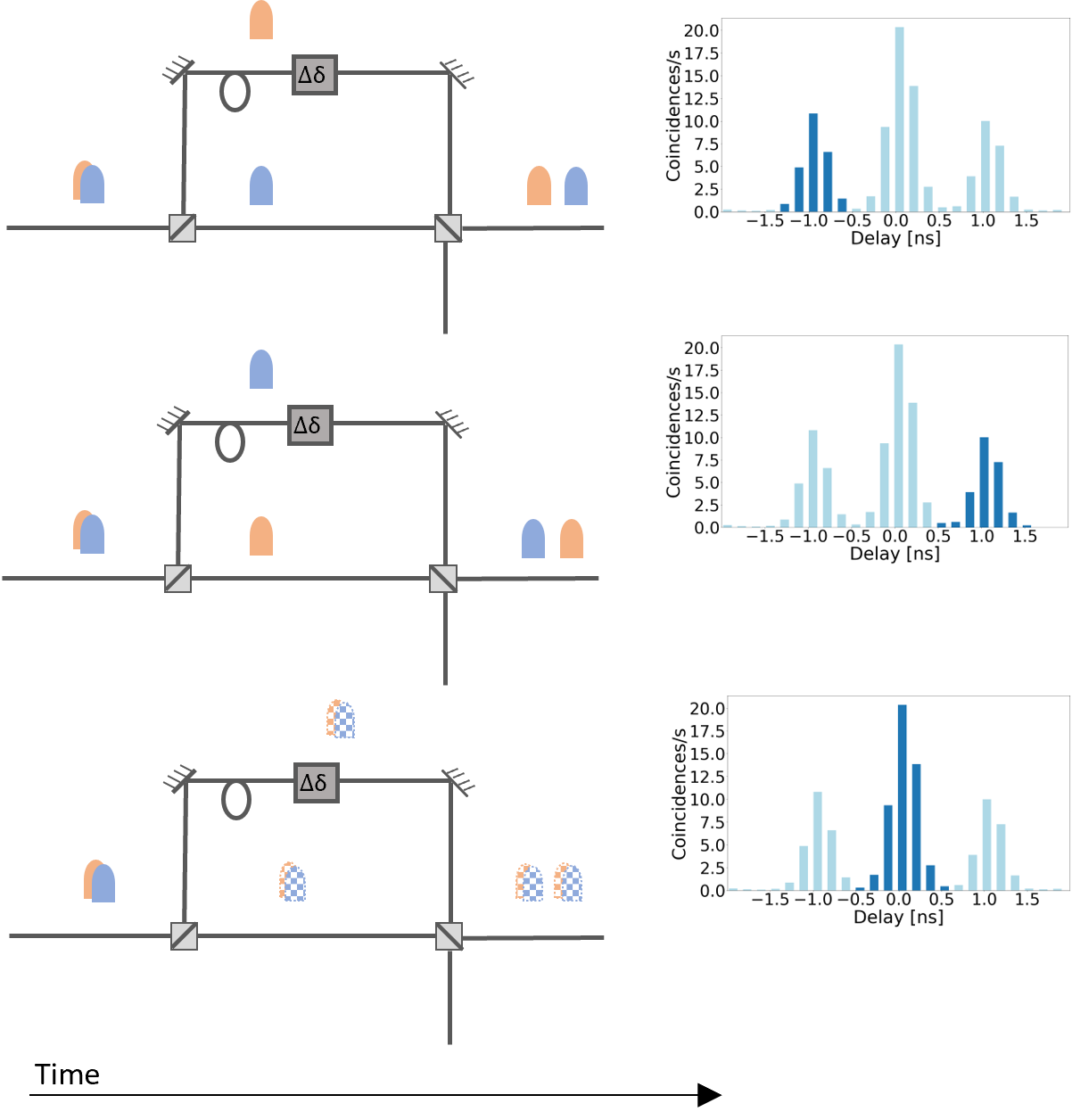}
\caption{The three possibilities for the path of the two photons originating from one time-bin are shown with the corresponding coincidence peaks highlighted in the delay histogram. The two peaks to the left and the right of the center peak show non entangled states, whereas the center peak on the bottom occurs for pairs taking the same path but can originate from sequential time-bins. The interference of the two probability waves then leads to entanglement.}
\label{fig:timebinMZI}
\end{figure}

In Fig. \ref{fig:timebinMZI} the three possible photon paths are visualized, when measuring the coincidences with a subsequent beamsplitter we see a histogram like the ones in Fig. \ref{fig:timebinMZI} with the corresponding peaks to the photon paths highlighted in dark blue. For further analysis we postselect the middle peak of the coincidences.
In a pulse train every time-bin could potentially contain a superposition of the current and the preceding incoming time-bin and its probability of carrying a photon pair, except the last and the first time-bin since they do not have adjacent time-bins. 
The post-selected quantum state from eq. \ref{eq:timebintransform} including adjacent time-bins (see eq. \ref{eq:spdcstate}) can now be written as (for simplicity the following eq. is not normalized)

\begin{equation}
\begin{split}
    \ket{\psi_{MZI}}=\ket{1,1}+\sum^n_{j=2}( \ket{j,j}_{l} \epsilon^{i\cdot\phi_j}+ \ket{j,j}_{e} \epsilon^{i\cdot \phi_{j-1} + 2\delta}) \\+\ket{n+1,n+1} \epsilon^{i\cdot \phi_{n+1} + 2\delta}
    \end{split}
\end{equation}

the subscripts e and l are referring to early (taking the long path) and late (taking the short path) incoming time bins.

\begin{figure*}[t]
\centering
\includegraphics[width=0.99\linewidth]{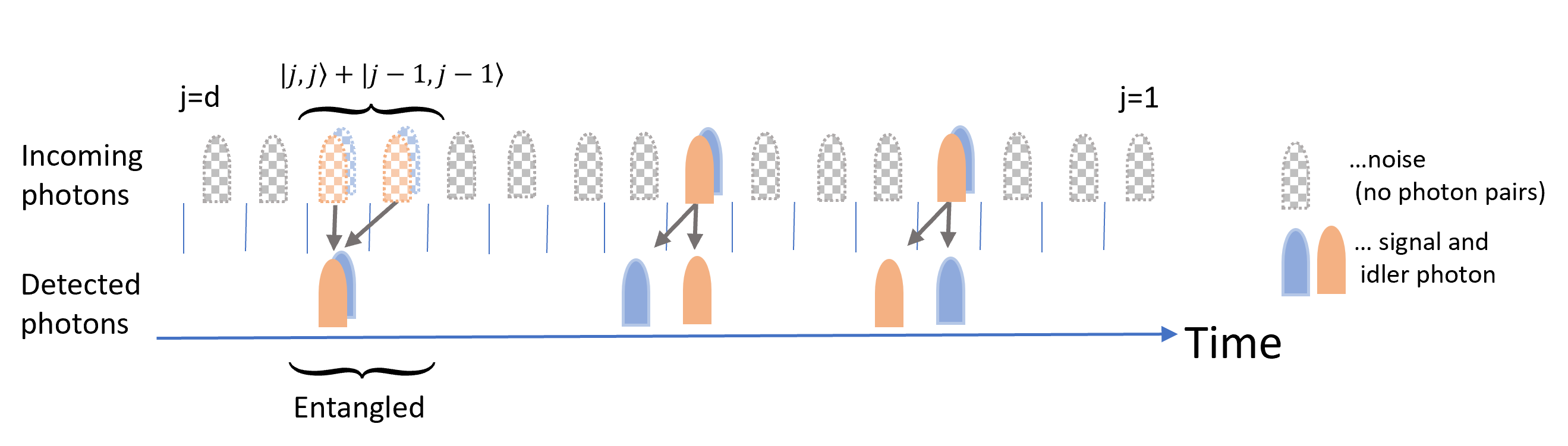}

\caption{The originating time-bins of signal and idler photons (blue and orange), have three different ways they can take in the MZI delay line, for the case were the photon pair ends up in the same time-bin after the MZI (found in the center peak) the photons are time-bin entangled.}
\label{fig:TimebinSequence}
\end{figure*}

The quantum state reduces to a two dimensional entangled state if we ignore the first and last time-bin. The remaining terms describe a coincidence detection in one time-bin that originates from a superposition of two adjacent input time-bins (see figure \ref{fig:TimebinSequence}).\\

In figure \ref{fig:TimebinSequence} the three possibilities for the time-bins is shown with the entangled state on the very left side, with the corresponding term in the resulting quantum state written above. It can be clearly seen that only the state with the ambiguity for the originating time-bin can be used for entanglement distribution.

Since the probability of a pulse train containing more than one pair within the laser coherence time is low the probability of having multi-partite states is negligible. Generating multi-partite states would only be feasible if there are two pairs in adjacent time-bins, which in the low pump regime is negligible \cite{de_riedmatten_tailoring_2004}.

To characterize the quality of entanglement the phase in the delay arm of the MZI can be changed to adjust the interference of the coincidence measurement and subsequently obtaining a visibility curve. In the coincidence histogram with the characteristic three peaks, only the middle peak is affected with a change of the phase in the MZI due to the interfering amplitude probabilities.\\

The probability for getting both pairs in the same resulting time-bin is the same as the probability for them being in different time-bins, as a consequence of this, the side peaks of the delay histogram are half as large as the middle bin.

\section{Entanglement distribution}

\begin{figure*}[h]
\centering
\includegraphics[width=0.99\linewidth]{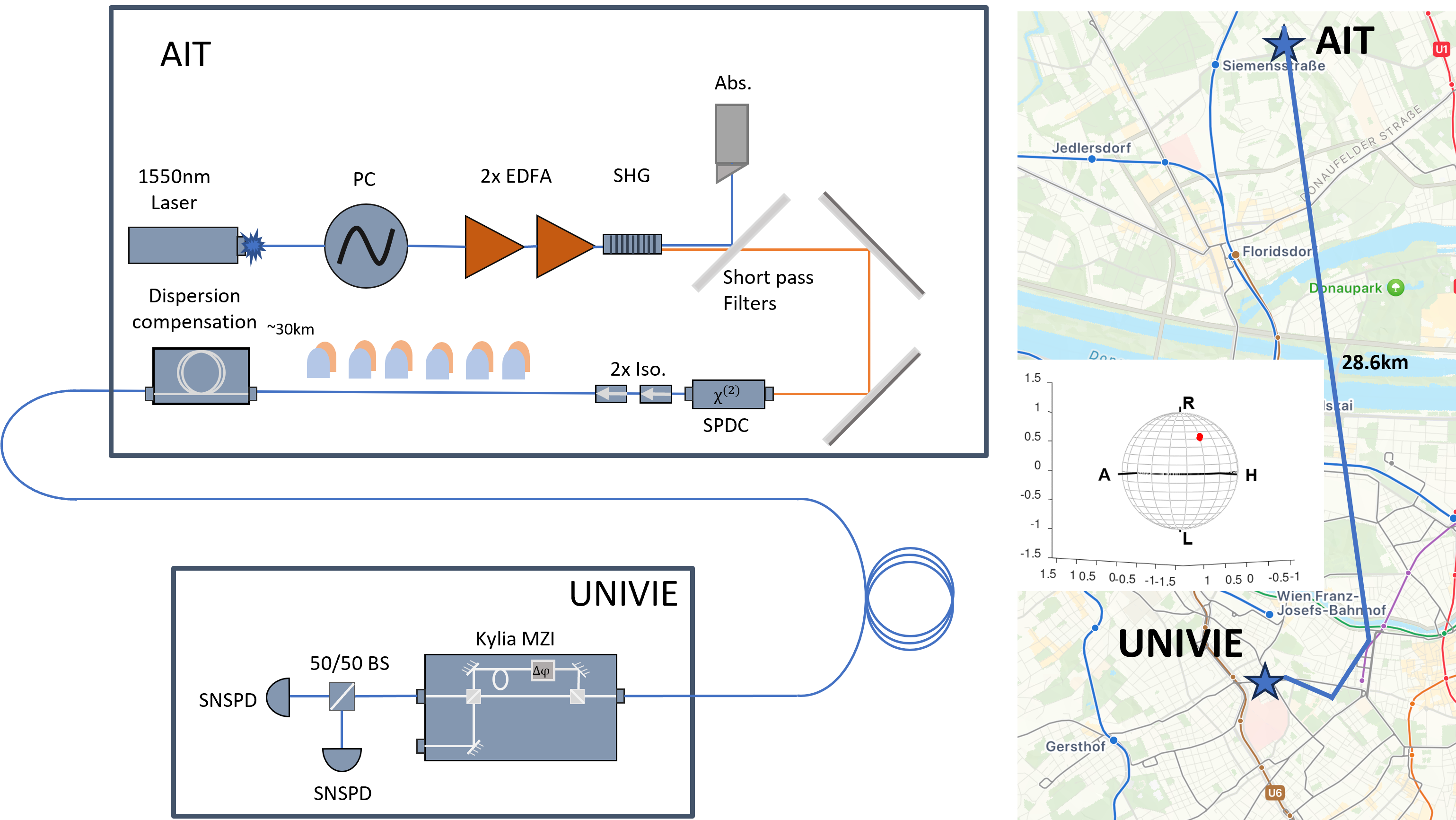}

\caption{On the left side a sketch of the experimental set-up is shown. The entanglement source is located  at the Austrian Institute of Technology (AIT) which is fiber linked to the University of Vienna (UNIVIE) where the setup for entanglement characterization is located. The right side portraits a map of Vienna (taken from Apple maps) with the approximate route the Fiber link is patched through, as well as the variation of the polarisation state on the Poincaré sphere over a time of around 2 hours. A dispersion compensation module was employed at AIT. PC: Pulse carver, SHG: Second Harmonic Generation, SPDC: Spontaneous Parametric Down Conversion, MZI: Mach-Zehnder Interferometer, BS: Beam Splitter, SNSPD: Superconducting Nanowire Single Photon Detector.}
\label{fig:FiberlinkSetup}
\end{figure*}

After a thorough characterization of all the components the sequential time-bin generation and distribution was tested via a 28.6km (9.5dB optical loss) fiber-link through the city of Vienna. The entanglement source was situated at the Austrian Institute of Technology (AIT) and connected to the Vienna fiber network where it was patched through to the University of Vienna (UNIVIE) where the receiver setup was placed (see Fig. \ref{fig:FiberlinkSetup}).

\begin{figure*}[h]
\centering
\includegraphics[width=0.99\linewidth]{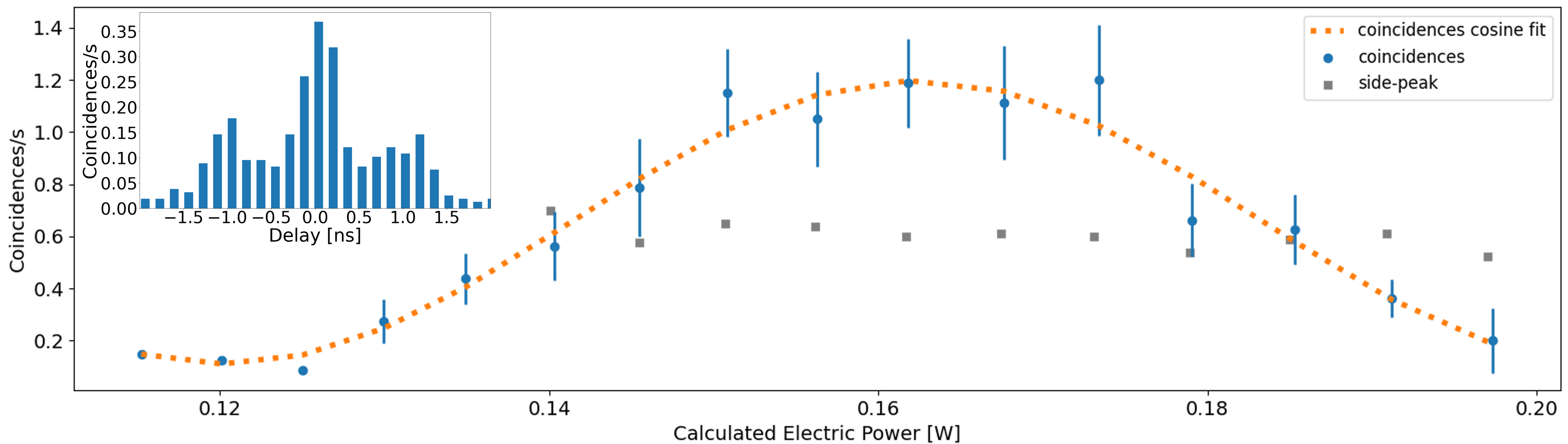}

\caption{Visibility curve measured after fiber-link of 28.6km length. The visibility was measured by an automated script changing the phase in the delay arm of the Kylia MZI, for each datapoint an idle time of around 5 minutes was implemented to let the temperature of the thermal element responsible for phase change stabilize together with 1 minute measurement time, this resulted in a measurement time of around 2 hours. The measurement showed an excellent corrected visibility of around 93\%. In the upper left part an inlay portraying an exemplary characteristic delay histogram of the three coincidence peaks corresponding to the different photon pair paths in the MZI delay line is shown. Since only the middle peak changes with the phase variation due to quantum interference, the average coincidences of the side-peaks are also displayed as grey data points in the main graph, showing the expected constant values.}
\label{fig:visibility}
\end{figure*}

The advantage of time-bin encoding compared to polarisation is that it makes the quality of the distributed entanglement immune with respect to polarisation drifts coming from the inherent birefrigence of the fibers. Over a time window of 2 hours the polarisation showed a good stability. This can be seen in the evolution of the polarisation state plotted on the Poincaré sphere in Fig. \ref{fig:visibility}. For a longer time however the polarisation does change considerably, mainly due to temperature variations a polarisation compensation algorithm has to be implemented to compensate for it \cite{neumann_continuous_2022}. In general the polarisation independence of the states provides considerable advantage for applications covering long distances.

When operating at GHz repetition rate, chromatic dispersion has to be carefully considered. The chromatic dispersion of standard single mode 1550nm optical fibers is on average around 18ps/(nm*km). Considering the width of the SPDC spectrum (3.5nm) and the 30km fiber link, chromatic dispersion produces a temporal broadening of the optical pulse from an original 150ps to about 2ns which is longer than the 1ns pulse separation. This broadening effectively washes out the time-bin entanglement if not compensated. To address this issue a commercial dispersion compensation module for 30km was introduced after the source at AIT. This module generates an inverse dispersion to cancel out the dispersion coming from the fiberlink, though adding a 2.9dB additional loss, resulting in a lower coincidence rate. The resulting FWHM of the coincidence peak at UNIVIE, as a consequence is at around 400ps which is larger than the measured FWHM at the back-to-back configuration (350ps) at AIT, which can be explained by remaining uncompensated dispersion.\\

After the MZI the entangled photons can be split with either a polarising beamsplitter (PBS) or a normal beam splitter (BS). Using a BS removes the polarisation dependence but introduces an additional 3dB loss. In case of relatively fast measurements a PBS is preferred for the sake of brightness optimization.

In figure \ref{fig:visibility}, the visibility curve measured at UNIVIE is displayed together with the characteristic delay histogram as an inlay in the upper-left side of the graph showing the three coincidence peaks portraying the possible combinations of signal and idler in the respective time-bins. The dispersion compensation is effective and the 3 peaks of the time-bin entanglement measurement are visible and well defined. Quantum interference fringes were measured by changing the phase in the delay arm of the interferometer. The phase in the MZI is changed by a thermal element in the delay arm, the dissipated power is shown on the x-axis of the graph. The curve is fitted with a cosine function displayed as the dotted orange line. The side-peaks were also measured and the measured coincidence value over the phase change is shown as the grey data points, which displays the expected constant value.

The coincidence window is set to 400ps, which corresponds to the estimated combined detection unit jitter at UNIVIE. The efficiency of the two SNSPDs is 80\%, with a jitter of about 150-200ps. The single counts were at a rate of around 20kc/s on top of background noise of approximately 15kc/s. The background noise increases the accidental coincidences hence lowering the uncorrected quantum visibility. From a preliminary characterization the background resulted from a broadband nature spanning from the visible to the telecom band and for the sake of simplicity of operation it has not been filtered out. A self-coded script varied the MZI electrical input over the shown range with around 5 minutes in between the actual measurements to let the set temperature stabilize. This resulted in a measurement time of around 2 hours for the visibility curve.  

The resulting raw visibility was at around 90\%, after correction via subtracting the accidentals the visibility increased to 93\%. The measured visibility showed a high degree of entanglement even after the 30km long fiber link, this is well above the 71\% CHSH limit for entanglement\cite{laudenbach_novel_2017}.
The SPDC spectrum, see Fig. \ref{fig:SPDCspectrum} has not been filtered, which is understood to limit the visibility. In comparison a back-to-back measurement (in the laboratory) showed a corrected visibility of 95 \% and a raw visibility of about 93\%. This indicates a high robustness of entanglement after distribution in the Vienna fiber network.

In a QKD network comprising two parties, the photon pairs can be split right after generation using, depending on the SPDC crystal, for example a PBS or a BS. The photons are then sent to their respective receiver which both are equipped with a MZI. Measuring the photons in their respective time-bins provides the necessary correlations for a secure quantum connection, enabling the two parties to exchange secret keys \cite{inoue_quantum_2005, kim_quantum_2022}.
One specific scheme for using time-bin entangled states for QKD requires both parties to have two detectors at their respective MZI outputs with detections in the two detectors corresponding to a 0 and a 1 bit. Here the side-peaks make up one basis (time bases) and the central peak the other basis (phase bases). Depending on which detectors click in each time bin at both receivers the two parties can infer a sifted key after sharing the basis in which the detection happened \cite{tittel_quantum_2000, brendel_pulsed_1999,fitzke_scalable_2022}. 

\section{Conclusion}
We showed the feasibility of distributing sequential time-bin entangled states in a metropolitan network. The results show high degree of entanglement even after a 30km fiber link, ready to be used in a QKD network with the inclusion of an additional MZI delay line and measurement equipment.
The entanglement generation setup at AIT and the analysis setup at UNIVIE were realized by using off-the-shelf components to facilitate a quickly developed deployment-ready entanglement source. To our knowledge this is the first time a commercial MZI delay line was used for a quantum application. This scheme can easily be scaled to higher repetition rates, assuming a low enough jitter of the detection equipment and suitable dispersion compensation. Our setup showed excellent visibility of around 93\% in the 30km fiber-link scenario which is well above the threshold for secure key generation \cite{scheidl_feasibility_2009} therefore offering usability for deployment as a QKD system (a second MZI receiver would be required). With the introduction of a polarisation compensation algorithm and a PBS instead of a BS after the MZI delay line losses can be reduced by 3dB which would result in around a 4 times increase in coincidences.
To increase visibility further investigation into the channel background noise is needed. We speculate that the current noise level originates from crosstalk in the fiber network.
In future projects we will be investigating the possibility of integrating the photon pair source as well as the MZI delay line on photonic chips \cite{ono_quantum_2024,purakayastha_-chip_2023,samara_high-rate_2019}.

With this scheme higher dimensional entangled states can also be generated by using an interferometer with multiple delay lines matched to the pulse time differences. This enables the probability wave to spread over more than two time-bins offering a higher dimensional entangled state, allowing for more complex applications.

\section*{Acknowledgments}

This work was supported by the Austrian FFG Agency (FAQT project, grant n° 884456). The authors would like to thank Lee Rozema and Michael Antesberger for support while performing the measurements at the University of Vienna.

\printbibliography

@article{scheidl_feasibility_2009,
	title = {Feasibility of 300 km quantum key distribution with entangled states},
	volume = {11},
	issn = {1367-2630},
	url = {https://iopscience.iop.org/article/10.1088/1367-2630/11/8/085002},
	doi = {10.1088/1367-2630/11/8/085002},
	pages = {085002},
	number = {8},
	journaltitle = {New Journal of Physics},
	shortjournal = {New J. Phys.},
	author = {Scheidl, Thomas and Ursin, Rupert and Fedrizzi, Alessandro and Ramelow, Sven and Ma, Xiao-Song and Herbst, Thomas and Prevedel, Robert and Ratschbacher, Lothar and Kofler, Johannes and Jennewein, Thomas and Zeilinger, Anton},
	urldate = {2024-06-05},
	date = {2009-08-04},
	langid = {english},
}

@article{neumann_continuous_2022,
	title = {Continuous entanglement distribution over a transnational 248 km fiber link},
	volume = {13},
	issn = {2041-1723},
	url = {https://www.nature.com/articles/s41467-022-33919-0},
	doi = {10.1038/s41467-022-33919-0},
	pages = {6134},
	number = {1},
	journaltitle = {Nature Communications},
	shortjournal = {Nat Commun},
	author = {Neumann, Sebastian Philipp and Buchner, Alexander and Bulla, Lukas and Bohmann, Martin and Ursin, Rupert},
	urldate = {2023-06-15},
	date = {2022-10-17},
	langid = {english},
	note = {Number: 1},
}

@article{bennett_quantum_1992,
	title = {Quantum cryptography without Bell's theorem},
	volume = {68},
	url = {https://link.aps.org/doi/10.1103/PhysRevLett.68.557},
	doi = {10.1103/PhysRevLett.68.557},
	pages = {557--559},
	number = {5},
	journaltitle = {Physical Review Letters},
	shortjournal = {Phys. Rev. Lett.},
	author = {Bennett, Charles H. and Brassard, Gilles and Mermin, N. David},
	urldate = {2024-10-21},
	date = {1992-02-03},
	note = {Publisher: American Physical Society},
}

@article{inoue_quantum_2005,
	title = {Quantum key distribution using a series of quantum correlated photon pairs},
	volume = {71},
	rights = {http://link.aps.org/licenses/aps-default-license},
	issn = {1050-2947, 1094-1622},
	url = {https://link.aps.org/doi/10.1103/PhysRevA.71.032301},
	doi = {10.1103/PhysRevA.71.032301},
	pages = {032301},
	number = {3},
	journaltitle = {Physical Review A},
	shortjournal = {Phys. Rev. A},
	author = {Inoue, Kyo},
	urldate = {2024-11-19},
	date = {2005-03-03},
	langid = {english},
}

@article{de_riedmatten_tailoring_2004,
	title = {Tailoring photonic entanglement in high-dimensional Hilbert spaces},
	volume = {69},
	rights = {http://link.aps.org/licenses/aps-default-license},
	issn = {1050-2947, 1094-1622},
	url = {https://link.aps.org/doi/10.1103/PhysRevA.69.050304},
	doi = {10.1103/PhysRevA.69.050304},
	pages = {050304},
	number = {5},
	journaltitle = {Physical Review A},
	shortjournal = {Phys. Rev. A},
	author = {De Riedmatten, Hugues and Marcikic, Ivan and Scarani, Valerio and Tittel, Wolfgang and Zbinden, Hugo and Gisin, Nicolas},
	urldate = {2024-11-19},
	date = {2004-05-18},
	langid = {english},
}

@inproceedings{zeiger_ps-pulse_2019,
	title = {A ps-pulse laser for ultrafast entanglement generation at 42.66 {GHz} repetition rate},
	rights = {© 2019 {IEEE}},
	url = {https://opg.optica.org/abstract.cfm?uri=EQEC-2019-eb_p_36},
	eventtitle = {European Quantum Electronics Conference},
	pages = {eb\_p\_36},
	booktitle = {2019 Conference on Lasers and Electro-Optics Europe and European Quantum Electronics Conference (2019), paper eb\_p\_36},
	publisher = {Optica Publishing Group},
	author = {Zeiger, Sophie and Laudenbach, Fabian and Schrenk, Bernhard and Hentschel, Michael and Hübel, Hannes},
	urldate = {2024-11-20},
	date = {2019-06-23},
}

@article{laudenbach_novel_2017,
	title = {A novel single-crystal \& single-pass source for polarisation- and colour-entangled photon pairs},
	volume = {7},
	rights = {2017 The Author(s)},
	issn = {2045-2322},
	url = {https://www.nature.com/articles/s41598-017-07781-w},
	doi = {10.1038/s41598-017-07781-w},
	pages = {7235},
	number = {1},
	journaltitle = {Scientific Reports},
	shortjournal = {Sci Rep},
	author = {Laudenbach, Fabian and Kalista, Sebastian and Hentschel, Michael and Walther, Philip and Hübel, Hannes},
	urldate = {2024-11-20},
	date = {2017-08-03},
	langid = {english},
	note = {Publisher: Nature Publishing Group},
}

@article{jin_efficient_2014,
	title = {Efficient generation of twin photons at telecom wavelengths with 2.5 {GHz} repetition-rate-tunable comb laser},
	volume = {4},
	rights = {2014 The Author(s)},
	issn = {2045-2322},
	url = {https://www.nature.com/articles/srep07468},
	doi = {10.1038/srep07468},
	pages = {7468},
	number = {1},
	journaltitle = {Scientific Reports},
	shortjournal = {Sci Rep},
	author = {Jin, Rui-Bo and Shimizu, Ryosuke and Morohashi, Isao and Wakui, Kentaro and Takeoka, Masahiro and Izumi, Shuro and Sakamoto, Takahide and Fujiwara, Mikio and Yamashita, Taro and Miki, Shigehito and Terai, Hirotaka and Wang, Zhen and Sasaki, Masahide},
	urldate = {2024-11-20},
	date = {2014-12-19},
	langid = {english},
	note = {Publisher: Nature Publishing Group},
}

@article{mancinelli_mid-infrared_2017,
	title = {Mid-infrared coincidence measurements on twin photons at room temperature},
	volume = {8},
	issn = {2041-1723},
	url = {https://www.ncbi.nlm.nih.gov/pmc/articles/PMC5440726/},
	doi = {10.1038/ncomms15184},
	pages = {15184},
	journaltitle = {Nature Communications},
	shortjournal = {Nat Commun},
	author = {Mancinelli, M. and Trenti, A. and Piccione, S. and Fontana, G. and Dam, J. S. and Tidemand-Lichtenberg, P. and Pedersen, C. and Pavesi, L.},
	urldate = {2024-11-20},
	date = {2017-05-15},
	pmid = {28504244},
	pmcid = {PMC5440726},
}

@article{thiel_time-bin_2024,
	title = {Time-bin entanglement at telecom wavelengths from a hybrid photonic integrated circuit},
	volume = {14},
	rights = {2024 The Author(s)},
	issn = {2045-2322},
	url = {https://www.nature.com/articles/s41598-024-60758-4},
	doi = {10.1038/s41598-024-60758-4},
	pages = {9990},
	number = {1},
	journaltitle = {Scientific Reports},
	shortjournal = {Sci Rep},
	author = {Thiel, Hannah and Jehle, Lennart and Chapman, Robert J. and Frick, Stefan and Conradi, Hauke and Kleinert, Moritz and Suchomel, Holger and Kamp, Martin and Höfling, Sven and Schneider, Christian and Keil, Norbert and Weihs, Gregor},
	urldate = {2024-11-20},
	date = {2024-05-01},
	langid = {english},
	note = {Publisher: Nature Publishing Group},
}

@article{anwar_entangled_2021,
	title = {Entangled photon-pair sources based on three-wave mixing in bulk crystals},
	volume = {92},
	issn = {0034-6748, 1089-7623},
	url = {https://pubs.aip.org/rsi/article/92/4/041101/961170/Entangled-photon-pair-sources-based-on-three-wave},
	doi = {10.1063/5.0023103},
	pages = {041101},
	number = {4},
	journaltitle = {Review of Scientific Instruments},
	author = {Anwar, Ali and Perumangatt, Chithrabhanu and Steinlechner, Fabian and Jennewein, Thomas and Ling, Alexander},
	urldate = {2024-11-20},
	date = {2021-04-01},
	langid = {english},
}

@article{ono_quantum_2024,
	title = {Quantum interference of pulsed time-bin entanglement generated from silicon ring resonator},
	volume = {14},
	rights = {2024 The Author(s)},
	issn = {2045-2322},
	url = {https://www.nature.com/articles/s41598-024-51311-4},
	doi = {10.1038/s41598-024-51311-4},
	pages = {1051},
	number = {1},
	journaltitle = {Scientific Reports},
	shortjournal = {Sci Rep},
	author = {Ono, Takafumi and Tsujimoto, Yoshiaki and Wakui, Kentaro and Fujiwara, Mikio},
	urldate = {2024-11-20},
	date = {2024-01-11},
	langid = {english},
	note = {Publisher: Nature Publishing Group},
}

@inproceedings{purakayastha_-chip_2023,
	location = {San Francisco, United States},
	title = {On-chip analysis of time-bin encoded photons},
	isbn = {978-1-5106-5997-1 978-1-5106-5998-8},
	url = {https://www.spiedigitallibrary.org/conference-proceedings-of-spie/12446/2649202/On-chip-analysis-of-time-bin-encoded-photons/10.1117/12.2649202.full},
	doi = {10.1117/12.2649202},
	eventtitle = {Quantum Computing, Communication, and Simulation {III}},
	pages = {86},
	booktitle = {Quantum Computing, Communication, and Simulation {III}},
	publisher = {{SPIE}},
	author = {Purakayastha, Ujaan and Nussbaum, Benjamin E. and Floyd, John C. and Evans, Chris C. and Hensley, Joel M. and Kwiat, Paul G.},
	editor = {Hemmer, Philip R. and Migdall, Alan L.},
	urldate = {2024-11-20},
	date = {2023-03-08},
	langid = {english},
}

@article{marcikic_time-bin_2002,
	title = {Time-bin entangled qubits for quantum communication created by femtosecond pulses},
	volume = {66},
	rights = {http://link.aps.org/licenses/aps-default-license},
	issn = {1050-2947, 1094-1622},
	url = {https://link.aps.org/doi/10.1103/PhysRevA.66.062308},
	doi = {10.1103/PhysRevA.66.062308},
	pages = {062308},
	number = {6},
	journaltitle = {Physical Review A},
	shortjournal = {Phys. Rev. A},
	author = {Marcikic, I. and De Riedmatten, H. and Tittel, W. and Scarani, V. and Zbinden, H. and Gisin, N.},
	urldate = {2024-11-20},
	date = {2002-12-10},
	langid = {english},
}

@article{kim_quantum_2022,
	title = {Quantum communication with time-bin entanglement over a wavelength-multiplexed fiber network},
	volume = {7},
	issn = {2378-0967},
	url = {https://pubs.aip.org/app/article/7/1/016106/2835124/Quantum-communication-with-time-bin-entanglement},
	doi = {10.1063/5.0073040},
	pages = {016106},
	number = {1},
	journaltitle = {{APL} Photonics},
	author = {Kim, Jin-Hun and Chae, Jin-Woo and Jeong, Youn-Chang and Kim, Yoon-Ho},
	urldate = {2024-11-20},
	date = {2022-01-01},
	langid = {english},
}

@article{samara_high-rate_2019,
	title = {High-rate photon pairs and sequential Time-Bin entanglement with Si$_{\textrm{3}}$ N$_{\textrm{4}}$ microring resonators},
	volume = {27},
	issn = {1094-4087},
	url = {https://opg.optica.org/abstract.cfm?URI=oe-27-14-19309},
	doi = {10.1364/OE.27.019309},
	pages = {19309},
	number = {14},
	journaltitle = {Optics Express},
	shortjournal = {Opt. Express},
	author = {Samara, Farid and Martin, Anthony and Autebert, Claire and Karpov, Maxim and Kippenberg, Tobias J. and Zbinden, Hugo and Thew, Rob},
	urldate = {2024-11-20},
	date = {2019-07-08},
	langid = {english},
}

@misc{riedmatten_creating_2002,
	title = {Creating high dimensional time-bin entanglement using mode-locked lasers},
	url = {http://arxiv.org/abs/quant-ph/0204165},
	doi = {10.48550/arXiv.quant-ph/0204165},
	number = {{arXiv}:quant-ph/0204165},
	publisher = {{arXiv}},
	author = {Riedmatten, Hugues de and Marcikic, Ivan and Zbinden, Hugo and Gisin, Nicolas},
	urldate = {2024-12-04},
	date = {2002-04-29},
	eprinttype = {arxiv},
	eprint = {quant-ph/0204165},
}

@article{fitzke_scalable_2022,
	title = {Scalable Network for Simultaneous Pairwise Quantum Key Distribution via Entanglement-Based Time-Bin Coding},
	volume = {3},
	url = {https://link.aps.org/doi/10.1103/PRXQuantum.3.020341},
	doi = {10.1103/PRXQuantum.3.020341},
	pages = {020341},
	number = {2},
	journaltitle = {{PRX} Quantum},
	shortjournal = {{PRX} Quantum},
	author = {Fitzke, Erik and Bialowons, Lucas and Dolejsky, Till and Tippmann, Maximilian and Nikiforov, Oleg and Walther, Thomas and Wissel, Felix and Gunkel, Matthias},
	urldate = {2024-12-13},
	date = {2022-05-24},
	note = {Publisher: American Physical Society},
}

@article{brendel_pulsed_1999,
	title = {Pulsed energy-time entangled twin-photon source for quantum communication},
	volume = {82},
	issn = {0031-9007, 1079-7114},
	url = {http://arxiv.org/abs/quant-ph/9809034},
	doi = {10.1103/PhysRevLett.82.2594},
	pages = {2594--2597},
	number = {12},
	journaltitle = {Physical Review Letters},
	shortjournal = {Phys. Rev. Lett.},
	author = {Brendel, J. and Gisin, N. and Tittel, W. and Zbinden, H.},
	urldate = {2024-12-16},
	date = {1999-03-22},
	langid = {english},
	eprinttype = {arxiv},
	eprint = {quant-ph/9809034},
}

@article{tittel_quantum_2000,
	title = {Quantum Cryptography Using Entangled Photons in Energy-Time Bell States},
	volume = {84},
	rights = {http://link.aps.org/licenses/aps-default-license},
	issn = {0031-9007, 1079-7114},
	url = {https://link.aps.org/doi/10.1103/PhysRevLett.84.4737},
	doi = {10.1103/PhysRevLett.84.4737},
	pages = {4737--4740},
	number = {20},
	journaltitle = {Physical Review Letters},
	shortjournal = {Phys. Rev. Lett.},
	author = {Tittel, W. and Brendel, J. and Zbinden, H. and Gisin, N.},
	urldate = {2024-12-16},
	date = {2000-05-15},
	langid = {english},
}

@article{bennett_quantum_2014,
	title = {Quantum cryptography: Public key distribution and coin tossing},
	volume = {560},
	issn = {03043975},
	url = {https://linkinghub.elsevier.com/retrieve/pii/S0304397514004241},
	doi = {10.1016/j.tcs.2014.05.025},
	shorttitle = {Quantum cryptography},
	pages = {7--11},
	journaltitle = {Theoretical Computer Science},
	shortjournal = {Theoretical Computer Science},
	author = {Bennett, Charles H. and Brassard, Gilles},
	urldate = {2024-12-16},
	date = {2014-12},
	langid = {english},
}

\end{document}